# State-of-the-Art Methods for Exposure-Health Studies: results from the Exposome Data Challenge Event

Léa Maitre⨱, Jean-Baptiste Guimbaud, Charline Warembourg, Nuria Güil-Oumrait, The Exposome Data Challenge Participant Consortium, Paula Marcela Petrone, Marc Chadeau-Hyam, Martine Vrijheid, Juan R. Gonzalez*, Xavier Basagaña*

*shared senior authorship

⨱ Corresponding author: lea.maitre@isglobal.org

## Abstract

The exposome recognizes that individuals are exposed simultaneously to a multitude of different environmental factors and takes a holistic approach to the discovery of etiological factors for disease. However, challenges arise when trying to quantify the health effects of complex exposure mixtures. Analytical challenges include dealing with high dimensionality, studying the combined effects of these exposures and their interactions, integrating causal pathways, and integrating high-throughput omics layers. To tackle these challenges, the Barcelona Institute for Global Health (ISGlobal) held a data challenge event open to researchers from all over the world and from all expertises. Analysts had a chance to compete and apply state-of-the-art methods on a common partially simulated exposome dataset (based on real case data from the HELIX project) with multiple correlated exposure variables ($P$>100 exposure variables) arising from general and personal environments at different time points, biological molecular data (multi-omics: DNA methylation, gene expression, proteins, metabolomics) and multiple clinical phenotypes in 1301 mother-child pairs. Most of the methods presented included feature selection or feature reduction to deal with the high dimensionality of the exposome dataset. Several approaches explicitly searched for combined effects of exposures and/or their interactions using linear index models or response surface methods, including Bayesian methods. Other methods dealt with the multi-omics dataset in mediation analyses using multiple-step approaches. Here we discuss features of the statistical models used and provide the data and codes used, so that analysts have examples of implementation and can learn how to use these methods. Overall, the exposome data challenge presented a unique opportunity for researchers from different disciplines to create and share state-of-the-art analytical methods, setting a new standard for open science in the exposome and environmental health field.

**Keywords:** exposome; statistical models; multi-omics; multiple exposures; Environmental exposures

# Background

The exposome, described as "the totality of human environmental exposures from conception onwards", recognizes that individuals are exposed simultaneously to a multitude of environmental factors and takes a holistic approach to the discovery of etiological factors for disease. The exposome's main advantage over traditional 'one-exposure-one-disease' study approaches is that it provides an unprecedented conceptual framework for the study of multiple environmental hazards (urban, chemical, lifestyle, social) and their combined effects. Given the increasing availability of complex environmental health data due to the emergence of new technologies (such as electronic health records, high throughput omics platforms, wearable sensors, etc), there is a need for more advanced statistical approaches that focus on complex mixtures of exposures.

To address the numerous challenges that come with the analysis of such complex data and to promote interdisciplinary collaboration between researchers from around the world, ISGlobal hosted a 3-day online data challenge in April 2021 entitled "the Exposome Data Challenge Event". This event gathered a widely diverse scientific audience of 307 participants, including environmental epidemiologists, biostatisticians and computational scientists, to discuss state-of-the-art statistical methods for studying exposome-health associations. First, participants were invited to submit an abstract describing their team, the challenge(s) and the method they would apply on a common partially simulated exposome dataset (based on real case data from the HELIX project). The planning committee selected a total of 25 abstracts out of 39 based on method clarity, novelty, relevance for the exposome field and challenges presented. Second, the selected participants were invited to apply their method on the dataset during a month leading to the event. Third, they presented at the event their method's statistical background, type of research question(s) it best addressed, and their results. At the end of the event the committee and the audience voted for the best presentations based on clarity, novelty and relevance. Finally, the participants made their code available on the [github account](#) (González, 2021) of the event.

Our main objective was to promote innovative statistical, data science, or other quantitative approaches to study the health effects of complex multi-dimensional exposures and high throughput omics measurements. To illustrate the methods discussed as part of this data challenge, it was important to create a publicly available [real case scenario dataset](#) (based on the [HELIX project database](#)). This dataset consisted of multiple correlated variables ($P>100$ exposure variables) arising from general and personal environments at different time points of early life, biological molecular data (multi-omics: DNA methylation, gene expression, proteins, metabolomics) and multiple clinical phenotypes. The population was drawn from the multi-center HELIX project, where the cohorts of origin represented the main confounding structure in the dataset (Maitre et al., 2018; Vrijheid et al., 2014)). The participants were offered an opportunity to test their statistical methods of choice addressing one or several key challenges: a) the high dimensionality of the data, b) combined effect of exposure or mixtures, c) the omics data integration and the d) causal structure in the exposome. Participants were encouraged to accommodate in their approaches some of the particularities of the data (e.g. multi-cohort, count responses, categorical and continuous exposure variables, exposures measured at two time points, etc). Visualization of the results was also a key point across all the challenges listed above.

This report outlines the approaches presented at the event, which represent useful computational, conceptual, and statistical models for analyzing high dimensional exposome datasets and health outcome associations. In collaboration with the event committee and the selected participants, we discuss the different techniques.

# Methods

**Data**

The exposome data provided for this challenge came from the HELIX subcohort database and were partially simulated. The HELIX study (Maitre et al., 2018; Vrijheid et al., 2014) represents a collaborative project across six established and ongoing longitudinal population-based birth cohort studies in six European countries (France, Greece, Lithuania, Norway, Spain, and the United Kingdom). From the 31 472 mother-child pairs included in the cohorts, a subcohort of 1301 mother-child pairs were followed up with measurements of biomarkers, omics signatures and health outcomes at 6-11 years of age. The data provided for the challenge came from this subsample, but it was partially simulated to respond to policies of data anonymization for privacy protection in the cohorts. In detail, for the set of health outcomes and exposures, we conducted the following process: for each participant, a total of 50 random variables (different for each subject) were converted into missing values and then imputed (in successive rounds of 10 variables at a time) with the method of chained equations. Thus, for each participant, some of the values in the provided dataset were real and some were simulated, in a way that precluded knowing what is real data and what is simulated. Omics data instead were kept intact but annotations of genes and metabolites were not provided. We provided an imputed dataset in which all missing values in the original data were imputed by the chained equations method. Exposure data were transformed (*e.g.* logarithmic, square root, tertiles) to achieve symmetric distributions with a homogenous range of values (**Annex 1**). The original raw data are available on request subject to ethical and legislative review. The "HELIX Data External Data Request Procedures" are available with the data inventory in this website: http://www.projecthelix.eu/data-inventory.

The datasets are available in the github repository of the challenge event and transcriptomic and Epigenomics through a FigShare account: https://figshare.com/account/home#/projects/98813. An overview of available data is shown in **Figure 1** and the complete codebook description is available in **Annex 1**. It includes more than two hundred environmental exposure variables, 13 covariates, six health outcomes [body mass index (BMI), asthma, birth weight, neurobehaviour, intelligence quotient (IQ)], and omics data (serum metabolome, urine metabolome, gene expression, methylation). Exposure, covariate and health outcome datasets contain both continuous and categorical values; health outcome data additionally included count data. Exposure and omic datasets both included highly correlated features (correlation >0.8).

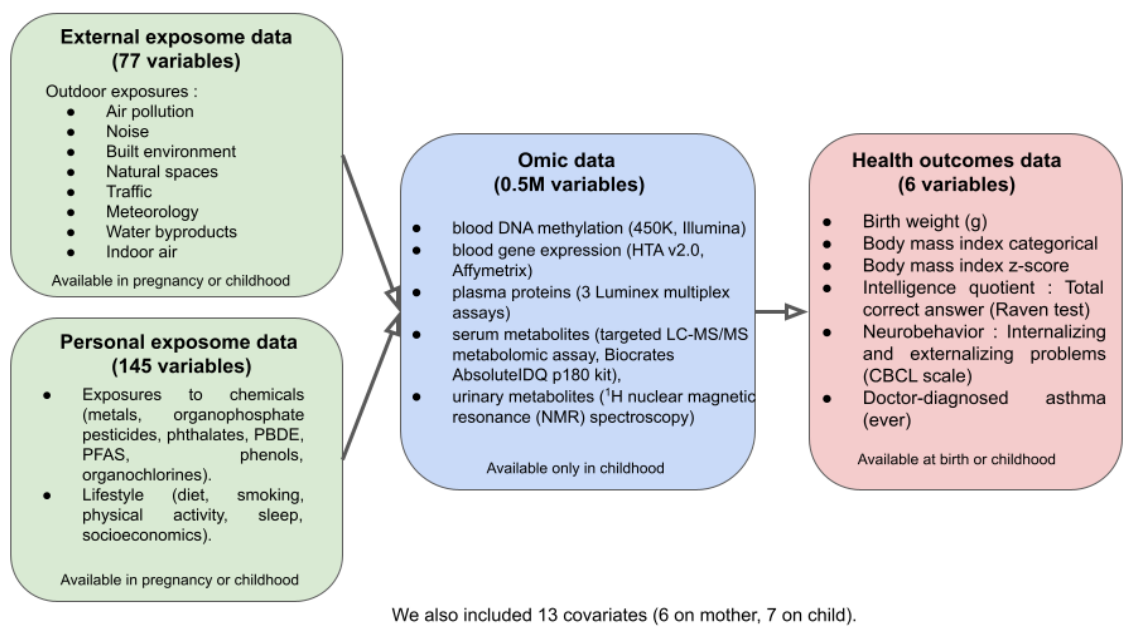

**Figure 1: Overview of the data available during the challenge.** The data were based on the HELIX project which collected exposome, omics and health data from six mother-child cohorts across Europe in 1301 participants (Maitre et al., 2018)

**Challenges**

The main challenges that were addressed during the workshop were as follows:

1/ The high dimensionality of exposure data and, more precisely, methods to reduce it while minimizing the loss of useful information regarding health associations and combined effects. The exposome dataset available during the challenge consisted of a large number of environmental exposure variables and a multi-omics dataset (approximately 0.5 M features), with a high inter-variable correlation, but a small sample size ($N$ = 1301), typical of exposome studies.

2/ The combined effect of exposure or mixtures. Researchers are interested in studying individual and combined effects of a large number of exposures together accounting for their potential interactions. Effects of environmental exposures may be small when taken individually, but their aggregation may lead to a significant alteration of the health outcome of interest, leading to cocktail effects that researchers want to investigate.

3/ Omics data integration. Omic data may be used, in addition to exposure data, in order to provide causal inference on the link between exposome and health. The challenge here was to incorporate one or several of the different omic layers available, with the purpose of finding patterns that can explain variations in one or more health outcomes and analysing how the exposome and the omes interact with regard to these outcomes. The method used in this context must be able to maximise omics data predictive power with very high dimensional data and small sample size ($N<<P$).

4/ Causal structures in exposure data. This challenge included: 1) how to incorporate *a priori* hypothesized causal relationships between the different exposures and one health outcome into the analysis, 2) the comparison of this *a priori* approach with *agnostic* analyses that would perform variable selection treating all exposures in the same way, 3) how one can answer a large number of causal questions referring to different exposures using causal inference techniques for high-dimensional data, 4) the incorporation of mediation analysis and high-dimensional mediation analysis.

Additional points of interest included visualisation techniques, the handling of the multicenter design of the study, the control for potential confounders that may have an effect on the health outcomes and need to be taken into account when studying associations with the exposomic features, and missing data in exposome datasets.

# Results

In this section, we summarise the statistical methods used by the participants in the Exposome Data Challenge listed in Table 1. All the codes developed by the participants to perform their analyses are available in the [GitHub repository](#) of the event. Briefly, most presentations focused on one health outcome out of the six available, mainly child BMI, and five used multiple outcomes. Seven presentations included categorical outcomes (*i.e.* Asthma, BMI and birth weight<2500gr), the rest focused on continuous outcomes only. Less than half of the methods included categorical exposure variables, focusing mainly on continuous chemical exposures. Eight presentations included omics data, including one with all omics layers. A separate results section is dedicated to the comparison of the findings across methods that focused on child BMI and chemical exposures.

**Table 1: Summary of all the presentations during the exposome data challenge.** Ge= Gene Expression, Me=DNA Methylation, Pr=Proteins, SM = Serum Metabolites, UM = Urine Metabolites, IQ = total correct answer (RAVEN test), Neurobehavior = Internalizing and externalising problems (CBCL scale).

| Presentation order | Authors names, University | Presentation Title | Method names | Link to slides | Link to papers | Omic data | Categorical exposures | Selected health outcome |
|---|---|---|---|---|---|---|---|---|
| 6 | Matthew Carli and David Wheeler, Virginia | Exposome Analysis with Bayesian Group Index Regression | Bayesian Group Index Regression | Link | (Wheeler et al., 2021) | SM | N | Asthma |

| | | | | | | | |
|---|---|---|---|---|---|---|---|
| | Commonwealth University | | | | | | |
| 5 | Shounak Chattopadhyay, Duke University | Synergistic Interaction Detection | Synergistic Interaction Detection | Link | *Article in preparation* | | N | Birth weight |
| 22 | Michele Peruzzi, Duke University | Multi-Outcome Meshed Gaussian Processes on Projected Inputs for Scalable Inference with Exposome Data | Meshed Gaussian Processes | Link | (Peruzzi et al., 2020) | | Y | Birth weight, BMI and related variables |
| 1 | Glen McGee, University of Waterloo | Quantifying Exposome-Health Associations with Bayesian Multiple Index Models | Bayesian Multiple Index Models | Link | (McGee et al., 2021) | | N | BMI z-score |
| 2 | Vishal Midya, Icahn School of Medicine at Mount Sinai | A novel penalized LASSO type Bayesian Weighted Quantile Sum Regression Approach for Exposome-outcome effect estimation | Bayesian Weighted Quantile Sum Regression | Link | (Colicino et al., 2020) | | N | BMI z-score |
| 13 | Fei Zou, University of North Carolina, Chapel Hill | Deep-Exposome: A Predictive and Interpretative Deep Neural Network Ensemble for Exposome Data | Improved Bootstrap Aggregating and PermFIT | Link | (Mi et al., 2021, n.d.) | | Y | All |
| 14 | Jean-Baptiste Guimbaud, Remy Cazabet, Léa Maitre, LIRIS-ISGlobal | Leveraging machine learning and explainable AI to better understand exposomic data | Multilayer Perceptron, xgboost, random forest, SVM, Elastic-net, SHAP | Link | NA | | Y | All |
| 10 | Ziyue Wang, National Institute of Environmental Health Sciences | Integrative Analysis and Visualization of Exposome and Transcriptome data | Differential expressed gene analysis (DEG) and Mediation analysis | Link | NA | GE | Y | Asthma |
| 15 | Alejandro Caceres, ISGlobal | Using causal random forest to determine exposure environments with high sexual dimorphisms | Causal random forests | Link | NA | Me, GE | Y | BMI z-score difference in boys and girls |
| 18 | Xiaotao Shen, Stanford University | Decoding unknown links between the exposome and health outcomes by multi-omics analysis | bi-directional mediation analysis. | Link | NA | GE, UM, SM, Pr | N | IQ, Neurobehavior, BMI z-score |
| 19 | Congrong Wang, Brigitte Reimann, Rossella Alfano, Hasselt University | Meet-in-the-middle meets multi-omics: a strategy to identify molecular signatures of environmental drivers of childhood BMI | Multi-omics Mediation Analysis | Link | NA | All | Y | BMI z-score |
| 20 | Daniela Zugna, University of Turin | Application of a novel method for mediation analysis in the exposome context | Mediation Analysis | Link | (Loh et al., 2020) | | Y | Birth weight (<2500gr) |
| 21 | Miao Yu, Icahn School of Medicine at Mount Sinai | Molecular Gatekeepers bridge the exposome and health | Molecular gatekeepers discovery | Link | (Yu et al., 2021) | SM | N | Asthma |
| 25 | Charlie Roscoe, Hari Iyer, Huichu Li, and Marcia Pescador Jimenez, Harvard University | Air pollution and childhood cognition: a g-computation approach to assess mediation by a mixture of metals | Causal mediation analysis and quantile g-computation | Link | (Keil et al., 2020) | | N | IQ |
| 9 | Nikos Stratakis, University of Southern California | Latent unknown clustering integrating multi-omics data (LUCID) with phenotypic traits | Unknown Clustering (LUCID) | Link | (Peng et al., 2020) | SM, UM | N | BMI z-score |

| | | | | | | | |
|---|---|---|---|---|---|---|---|
| 11 | Sejal Mistry and Ramkiran Gouripeddi, University of Utah | Clustering Exposure Trajectories to Classify Phenotypic Characteristics | clustering transitions on phenotypic characteristics | Link | NA | | N | BMI z-score |
| 12 | John Pearce, Medical University of South Carolina | Exposure Continuum Mapping for predicting health and disease in exposome studies | Exposure Continuum Mapping and Generalized Additive Models | Link | (Pearce et al., 2021) | | N but could | Birth weight |
| 3 | Ander Wilson, Colorado State University, Daniel Mork | Exposome Health Association Studies Using Bayesian Treed Distributed Lag Mixture Models | Treed Distributed Lag Mixture Models | Link | (Mork and Wilson, 2021) | | N but could | BMI z-score |
| 7 | Qiong Wu, University of Maryland, College Park | A new statistical graph model to systematically study associations between multivariate exposome data and multivariate metabolomics data | Bipartite Graph | Link | NA | SM | N | None |
| 4 | Chris Gennings, Icahn School of Medicine at Mount Sinai | Evaluating a Mixture Effect of Perinatal Environmental Exposures on Childhood BMI Using Weighted Quantile Sum (WQS) Regression | Weighted Quantile Sum Regression (WQS) | Link | (Carrico et al., 2015) | | N | BMI z-score |
| 8 | Jaime Benavides and Lawrence Chillrud, Columbia University | Pre- and postnatal urban exposure patterns and childhood neurobehavior | Principal Component Pursuit (PCP), Factor Analysis, GAM and LASSO | Link | (Gibson et al., 2021) | | N | Neurobehavior |
| 16 | Lucile Broséus and Paulina Jedynak, Université Grenoble Alpes | Searching for the risk factors for childhood overweight - A novel approach to identify the most relevant child BMI-associated exposures | Univariate Ordinal Logistic Regression and Multiple Correspondence Analysis (MCA) | Link | NA | | Y | BMI categorical |
| 17 | Carl Brunius, Chalmers University of Technology | Omics Modules for Exposome-Health Associations (OMEXA) | MUVR, GeneralizedLinear Models and Triplot | Link | (Shi et al., 2019) | All | Y | All |
| 23 | Sanjib Basu, Ruizhe Chen, Yu-Che Chung, Jiyeong Jang Mary Turyk and Hua-Yun Chen, University of Illinois at Chicago | Missingness pattern and exposure selection for mixed-type exposome data | COrrelation LeaRNing and exposure Selection (COLRNS) and A Test for Realized Missing Completely At Random | Link | NA | | Y | BMI z-score |
| 24 | Hua Yun Chen, University of Illinois at Chicago | Estimating the effects of exposome and their interactions | Explained variation (EV) in linear models | Link | NA | | Y | Birth weight, IQ |

1. **Approaches to deal with high dimensionality of the data**

Most analyses presented at the event, even those dealing only with environmental exposures (n>p), applied some sort of dimensionality reduction techniques, as summarized in Table 2. These techniques can be split in two categories: 1) feature extraction techniques (Khalid et al., 2014), which consist in computing derived variables that are functions of the original ones and have a smaller dimension, but retain/extract most of the information contained in the original feature space, and 2) feature selection techniques which consist in selecting a subset of the original variable set, while keeping most of the information contained in the whole variable space.

Among the feature extraction techniques that were used during the event, we can further define different types: 1) techniques that calculate summary measures (indices) based on the creation of weighted combinations of exposures that predict the health outcome; 2) feature projection techniques such as Principal Component Analysis (PCA) (Jolliffe, 1986), Factor Analysis (FA), Principal Component Pursuit (PCP) (Candes et al., 2009) or Multiple Correspondence Analysis (MCA) (Blasius and Greenacre, 2006) to represent the data in a low dimensional space; and 3) techniques that provide clustering of participants sharing a similar exposome profile, which could be predictive of the outcome, *i.e.* supervised, or not.

Feature selection approaches used during the event can also be divided into different types of selections: 1) based on correlation with the outcome of interest, using Pearson's correlation as a screening approach; 2) based on regularisation (LASSO, Elastic-net regression) by shrinking the less relevant features' coefficients; 3) based on feature importance which reflects the impact of a given feature on the model predictions through permutation (Altmann et al., 2010), random forest (Breiman, 2001), or regression coefficients. Other statistical tests exist for features selection (ex: chi-squared test, ANOVA, etc..) but were not used during this challenge. Some analysts made an *a priori* selection, by choosing to focus only on a particular subset of exposures, *e.g.* lifestyle exposure, based on prior knowledge for causal models (25) or model abilities (continuous-only-exposures).

Ten studies implemented feature selection but not feature extraction, five implemented feature extraction but not feature selection, and nine applied both techniques (Table 2).

**Table 2. Method classification according to dimensionality reduction technique used in the exposome data challenge.** The number of the presentation corresponds to the list in Table 1. The ones with an * correspond to presentations which can fit in several categories.

|  |  | Feature selection | | | |
| --- | --- | --- | --- | --- | --- |
|  |  | None | By statistical tests *(correlation)* | Regularized regression *(LASSO, Elastic-net)* | Feature importance *(tree based, permutation based, regression coefficients etc..)* |
| **Feature extraction** | None | 12, 13, 15, 24 | 7, 18*, 21 | 3, 5, 20 | 14, 16*, 17, 18* |
|  | Indices *(linear regression, weighted quantiles, risk scores)* | 25 |  | 1, 2, 10 | 1, 2, 4 |
|  | Feature projection *(PCA/FA/PCP/MCA)* | 11*,22 |  | 8 | 19 |
|  | Clusters | 9,11*, 12, | 23 |  |  |

Abbreviations: FA, Factor Analysis; PCA, Principal Component Analysis; PCP, Principal Component Pursuit; MCA, Multiple Correspondence Analysis

2. **Approaches to study the combined effect of exposures on health**

**2.1 Approaches explicitly searching for combined effects of exposures and/or their interactions**

Several methods were presented to capture the effect of exposure mixtures. Most of them can be classified as linear index models or response surface methods. Linear index models generate new variables (usually called indices) that are weighted averages of the original exposures, and regress those indices against the health outcome. Response surface methods fit a complex high-dimensional surface to the data, and are thus able to capture complex non-linearities and interactions.

In the group of index models, Gennings (4) et al. presented the weighted quantile sum regression (WQSR) method (Carrico et al., 2015), its history and recent extensions. WQSR builds a new index, which is a weighted average of the initial exposures (previously categorised into quantiles as a way to standardise the data and prevent the effect of influential observations). The new index is regressed against a health outcome, producing a single regression coefficient. The weights to build the index, which incorporate directionality constraints (e.g., all variables are expected to produce negative effects on the health outcome), are estimated simultaneously with the regression coefficients. This technique assumes additive effects of the different pollutants. Some presented extensions included models to produce strata-specific weights and regression coefficients, combining two indices (one for each directionality) in the same model, or using resampling to improve the properties of the method.

Carli (6) et al. presented the application of Bayesian Group Index Regression (BGIR), a Bayesian equivalent to WQSR that does not use directionality constraints and allows multiple indices (based on groups of exposures) in the same model (Wheeler et al., 2021). The method R package *BayesGWQS* is available [here](). In the application, they included several indices according to exposure families, and each exposure family could have two indices if the family contained both exposures that were positively and negatively correlated with the outcome. They conducted analyses separately by cohort and included serum metabolomics data as an additional group of exposures. Midya (2) et al. presented the application of LASSO-type Bayesian Weighted Quantile Sum Regression (LBWQSR), which is similar to BGIR, but introducing LASSO and Elastic net penalties to prevent overfitting (Xu and Ghosh, 2015).

In the group of response surface methods, Chattopadhyay (5) et al. presented a method to search for two-way non-linear interactions. Two-dimensional splines were used to capture the shape of the association for all pairs of continuous exposures. A Bayesian paradigm was used, with priors that allow shrinking terms to zero in the absence of interaction. Prior information on the direction of the interaction (synergistic vs. antagonistic) was also incorporated.

Mork et al. (3) presented the application of Treed Distributed Lag Mixture Models (TDLMM) (Mork and Wilson, 2021). This method uses a Bayesian additive regression trees style model that performs exposure selection of main effects and two-way interactions, and incorporates the repeated exposure measurements available at two time points. The model performs hierarchical variable selection (interactions are only included if both main effects are included), performs shrinkage of regression coefficients, and performs dimension reductions by averaging over multiple time points when there is no evidence that the association varies over time.

Peruzzi *et al.* (22) presented the application of Multi-outcome Meshed Gaussian Processes on Projected Inputs (PIMGP). This method adapts Meshed Gaussian Processes, a method from the geostatistical literature which normally works with bidimensional inputs, for use in higher dimensional input spaces: after projecting the inputs onto a lower dimensional subspace (using, *e.g.* PCA), PIMGP use common GP kernels and lead to much faster performance relative to standard GPs or Bayesian Kernel Machine Regression (BKMR), especially with big dataframes (high *N*). The modeling framework was very flexible, allowing multiple outcome variables, missing covariates and covariates measured with error.

McGee et al. (1) presented the application of Bayesian Multiple Index Models (BMIM) (McGee et al., 2021). This approach combines the dimension reduction and interpretability of linear index models (such as WQSR and BGIR) and flexible exposure-response modelling of response surface methods (such as BKMR and PIMGP). With BMIM, the original exposures are reduced to a set of indices, as in BGIR. The approach simultaneously estimates the index component weights (with variable selection) and a potentially complex, high-dimensional exposure-response relationship between the indices and health outcome. Thus, it allows non-linear effects of the indices and interactions between indices. This presentation won one of two exposome data challenge prizes.

## 2.2. Approaches using Machine Learning to maximize prediction performance

Guimbaud (14) et al. linked the exposome with several health outcomes using several machine learning methods, namely multilayer perceptron, random forest, XGboost, support-vector machines (SVM), and Elastic net. They compared the prediction performance of the different techniques and used explainable AI by calculating SHAP (SHapley Additive exPlanations) values (Lundberg and Lee, 2017) to examine the impact of each exposure in the resulting models and their interactions with regards to the health outcomes. Zou (13) et al. conducted similar analyses, using in this case a deep neural network ensemble model, which was compared in terms of prediction accuracy to LASSO, SVM and random forest. Instead of SHAP values, they calculated a permutation-based feature importance test. Broséus (16) et al. studied the relationship between the exposome and child's BMI using a multistep approach. First, they ranked predictors by feature importance using multivariate ordinal random forests. Based on this metric, they used an arbitrary threshold to select the most pertinent exposures and, among these, they performed an analysis of exposure associations using: 1) an ordinal logistic regression model to obtain effect estimates and direction of association and 2) an MCA to obtain a graphical view of the clustering of exposures, adapted to categorical exposures. Finally, Yun Chen et al. (24) proposed a measure of Explained Variation to assess the performance of models, which can be estimated accurately without estimating the individual regression coefficients. Looking at explained variation can provide interesting insights, such as confidence intervals, the explanatory power of different exposure families or of interactions terms.

## 2.3 Multi-stage approaches for combined effects

Multi-stage modelling approaches were applied to model patterns in the exposome data prior to examining associations with the outcome. For example, Pearce et al. (12) linked the exposome to birth weight by applying the framework defined as Exposure Continuum Mapping (ECM). An exposure continuum map is a spatially organized map of exposome features that places similar exposure profiles close to each other and different ones are further apart. It is built in two steps: first they build a low dimensional (2D) representation of the data using Kohonen self-organising maps (Kohonen, 1982) in order to identify exposure profiles. Then, using information from this organised map, they used a Generalized Additive Model (GAM) to build a 3D exposure-response function that allows examination of a total mixture effect.

Benavides et al. (8) linked the exposome with neurobehavior using a strategy that involved reducing the exposome via PCP (Gibson et al., 2021) and FA, in order to identify both consistent and unique exposure patterns, and then regress these lower dimensional patterns to the health outcome using generalized additive models (for the consistent patterns) and LASSO (for the unique patterns). Mistry (11) et al. linked the exposome to obesity. They used PCA and k-means clustering to identify exposure profiles in both the prenatal and postnatal periods, and then used logistic regression to assess the risk of obesity as a function of the transitions of individuals between prenatal and postnatal exposure clusters. Basu (23) et al. designed an iterative algorithm (COLRNS) that creates clusters of correlated exposures and then performs variable selection within the clusters to predict the health outcome while minimizing the error of the model. This method can also handle missing data.

### 3. Studies using omics data to improve inference on the link between exposome and health

Several studies used one or more omics datasets as intermediate layers and conducted some kind of mediation or meet-in-the-middle analyses. Wang (Ziyue) et al. (10) studied the link between the exposome and asthma using transcriptome as an intermediate layer. In particular, they used a combination of 1) differential gene expression and gene set enrichment for asthma, 2) exposure selection in a model for asthma *via* Elastic net and calculation of exposure risk scores, and 3) high-dimensional mediation analysis. Shen et al. (18) also conducted mediation analysis, in this case using several omics datasets (transcriptome, proteome, serum/urine metabolome) as potential mediators in the relationship between the exposome and several health outcomes (IQ, behavior, BMI). All models were fitted with linear mixed models and they used BH correction to correct for multiple comparisons. This presentation received the committee prize for integrating all the high dimensional omics and multiple outcomes while using informative visualisation of the results. Wang (Congrong) et al. (19) conducted a causal mediation analysis using multi-omics layers (transcriptome, proteome, serum/urine metabolome) as potential mediators of the relationship between the exposome and BMI. In this case, they used multi-omics factor analysis to reduce dimensionality of the omics layers and factor analysis to reduce the dimensionality of the exposome.

Yu et al. (21) presented Gatekeepers (Yu et al., 2021) a new theory to assess exposure-metabolites associations. They identified some Gatekeepers in the data using the Pearson correlation and then studied their associations with asthma. Gatekeepers are metabolites associated with both exposures and other metabolites. They are presented as the bridge between the exposome and the metabolome. Brunius et al. (17) also implemented a meet-in-the middle approach in which they used the proteome, serum metabolites, urine metabolites, gene expression and methylation as middle layers between the exposome and health outcomes. In their analytical pipeline, Omics Modules for Exposome Health Associations (OMEXA), they used machine learning (MUVR Multivariate Methods with Unbiased Variable Selection, a predictive machine learning algorithm using an embedded recursive feature elimination mechanism within a repeated double cross validation procedure) to select the exposure variables related to the phenotypes available and partial correlation to further refine the selected list of exposures, adjusting for covariates. Then, they reused the same pipeline to select omic variables related to the selected exposures; and finally, they implemented a generalized linear model to link the selected omics with the phenotypes. They visualized the final results with triplots by projecting exposures, omics and phenotypes into two principal components.

Zhao (Stratakis) et al. (9) studied the link between organochlorines and BMI, using proteomics and urine and serum metabolites as intermediate layers. They used the latent unknown clusters (LUCID) method, which found latent subgroups of subjects characterized by having at the same time distinguished BMI, distinguished omics profiles and distinguished exposure. The process includes variable selection with LASSO and results were visualized with a Sankey diagram.

Finally, Wu et al. (7) linked the exposome with metabolomics, without including the health phenotypes. They used a bipartite graph model to represent pairwise associations between exposures and metabolites. From this graph they extracted subgraphs of concentrated most significant negative or positive association blocks (that were assessed using the pairwise Pearson correlation coefficients).

  4.  **Causal inference**

Some researchers studied causal relationships between environmental exposure and health. Roscoe et al. (25) studied the relationship between air pollution and cognition, and used blood concentrations of metals (part of the exposome) as potential mediators. They used causal mediation analysis and quantile g-computation to assess mediated effects. Zugna et al. (20) also conducted mediation analyses, in this case they considered the exposome as a potential mediator in the association between socioeconomic position and birthweight. Thus, they studied a context with a high-dimensional mediator set, and their proposed analysis was based on interventional effects and penalized regression models (Loh et al., 2020).

Cáceres et al. (15) conducted an analysis trying to explain the differences in BMI between boys and girls (outcome variable). They inferred groups (clusters) of participants with specific exposomic profiles for which those differences were the highest. The underlying method used was a recent implementation (https://github.com/teff-package/teff) of random causal forests (Wager and Athey, 2017). They also looked for omics markers (transcriptomic and methylomic) that were associated with differences between the exposome clusters. This presentation won the popular vote of the challenge.

  5.  **Results on chemical pollutants and zBMI**

Most of the approaches which focused on child BMI confirmed previous HELIX publication results (Vrijheid et al., 2020) that childhood hexachlorobenzene (HCB) exposure is cross-sectionally associated with reduced childhood BMI z-score. Some studies also identified metals and PCBs (polychlorinated biphenyls, particularly PCB170) to be linked with BMI.

In a WQS regression of 38 prenatal and postnatal chemicals, Gennings et al. (4) reported a significant negative association between the mixture index and child BMI. Higher weights belonged to the postnatal exposures, with HCB being the chemical with the highest contribution. Conversely, Midya (2) et al. found in a penalized group mixture BWQSR that prenatal organochlorine compounds (OCs) and metals were positively associated with child BMI, whereas PCBs were negatively associated. Within each group, the chemicals with the highest weight were: HCB (for OCs), As, Cd and Co (for metals), and PCB170 (for PCBs).

Mork and Wilson (3) developed treed distributed lag mixture models with 56 prenatal and postnatal exposures and observed that the chemicals with the highest PIPs (near 1) were HCB, PCB170, DDE, and Mo. They also identified a strong interaction (PIP=1) between prenatal Mo and postnatal HCB. McGee et al. (1) grouped 150 exposures into 29 indices corresponding to exposure families and time of exposure (prenatal or postnatal) in a BMIM. The groups with the highest PIPs (>0.5) were postnatal OCs, postnatal metals, prenatal water DBPs and the postnatal built environment. Among these, OCs were strongly negatively associated with child BMI z-score, and HCB was the chemical with the strongest effect in the index, followed by PCB170 and DDE. No interactions were observed except for prenatal water DBPs and postnatal OCs. Consistently, Broséus et al. (16) also identified postnatal HCB as the chemical exposure with the highest importance in multivariate ordinal random forests. After combining lifestyle and chemical exposures in an MCA, they found that Cu was associated with an increased risk of childhood overweight.

Using the LUCID method, Yinqi Zhao et al. (9) identified protein signatures (IL-1beta, IL-6, insulin) giving insight into underlying mechanistic pathways of childhood obesity (eg., systemic inflammation, disturbed glucose metabolism). Finally, in multi-omics mediation analysis, Wang (Congrong) et al. (19) detected urine metabolites (e.g. phospholipids, TMAO, hippurate) that mediate the effect of maternal smoking and built environment on childhood BMI.

# Discussion

This event brought together researchers from various disciplines to work on a common challenge: exposome data analysis. It established an overview of state-of-the-art methods currently used in the field and paved the way to interesting discussions and exchange of ideas. The first point of interest was the wide use of dimensionality reduction methods, especially feature selection, to deal with multivariate exposomic data (even without omics). Feature extraction methods were less used because they usually complicate the interpretation of the results if we are interested in the effect of a particular exposure on health. However, during the challenge, some interesting methods tried to analyse groups of correlated exposures as a way to reduce the dimensionality of the input while keeping the results interpretable.

While there is still a use of linear methods (LASSO, ExWAS, Elastic-Net) to analyse exposome-health associations, there is a great interest in methods that can capture more complex non-linear relationships between the features and the health outcome. Among those methods, most of them were Bayesian methods that are able to approximate a wide class of functions while also giving credible intervals. The Bayesian paradigm is useful because it naturally penalizes complex models and it offers flexibility to incorporate a process of variable selection. It is worth noting that, even if models can capture complex high-dimensional surfaces, interpretation of such models is usually done using plots that show the effects of just one or two variables at a time. In fact, some studies suggest that structures based on four variables are at the limit of human ability to correctly process the information (Halford et al., 2005). In this data challenge, a few studies detected two-way interactions in the data based on Bayesian methods (synergistic interactions, multiple index-based or additive regression tree-based) or machine learning models combined with XAI techniques such as SHAP (Lundberg and Lee, 2017).

The limitations and complexity of interpreting high-dimensional surfaces are addressed by changing the research question to that of estimating a mixture effect associated with health outcomes using weighted indices. The detection of a mixture effect is more powerful based on a single degree of freedom test. Detection of non-negligible weights provides evidence of important components. Methods have been developed to estimate quadratic associations with the weighted index where the significance of the quadratic term may be used as a goodness of fit for an assumption of linearity. Of course, the limitation of an index approach has been that interactions cannot be detected. Recently addressed, multiple index models allow for interactions between indices and provide a compromise between single or additive index models and high-dimensional exposure-response surface methods (McGee et al., 2021).

Despite their great predictive power, machine learning techniques were until now set aside in the context of environmental health studies because of their lack of interpretability. Black box models are able to capture more complex information from the data (e.g., complex interactions, non linear relationships etc.) than linear, regression-based models, and this can be key to accurately estimate the impact of the exposome on health. Black box methods used during the challenge included ensemble methods (such as random forests, xgboost), neural networks and support-vector machines. The method most often used was by far random forests. One can improve on the interpretability of black box models in several ways, here we will discuss two of them. The feature importance metric was the most popular technique during the challenge and is also the most used in the exposome field. It can be computed in several ways depending on the model: 1) impurity based feature importance (Breiman, 2001) for a model based on decision trees 2) permutation based feature importance (Altmann et al., 2010) and shapley values based feature importance (Harris et al., 2021) can be both applied on any model (model agnostic). Another approach is to use partial dependence plots (Zhao and Hastie, 2021) that allow visualizing partial associations between variables. A related example is the application of a combination of super-learner and g-estimation to assess the association between chemical pollutants and cognitive function (Oulhote et al., 2019). Many of the machine learning methods used during the challenge were limited by the small sample size available ($N$=1301) and it is also worth mentioning that the data imputation could lead to spurious associations.

Considering the analysis of omic data, all studies performed some sort of dimensionality reduction before applying different statistical analysis. This is a limitation of most existing multi-omic data integration approaches since they have not been implemented to deal with large matrices. Therefore, development of new integrative methods/tools must consider efficient handling of large data sets (Subramanian et al., 2020). Most presentations studied the relationship between environmental exposures and health outcomes using omic data (single or multi-omics) as an intermediate layer in a mediation analysis fashion. Some other studies used Pearson correlation to study the relationships between omics and exposome data. Different combinations of statistical tools were proposed for omics data integration in analytical frameworks. These tools taken individually were not novel but their combination for integrating exposome, multi-omics and outcomes was strongly relevant and novel. We note that other methods previously used for exposome and omics data may also be of interest but were not presented during this event. These methods include dimension reduction derived from PCA such as partial least square (PLS) and its derivatives (sparse-PLS, group-PLS) (Chun and Keleş, 2010; Jain et al., 2018; Lenters et al., 2015), canonical-based methods or network analysis (Bessonneau et al., 2021).

Causality in exposome-health models were also assessed through mediation analysis including omics data, g-computation, and causal random forest. We note that none of the methods presented, addressed the challenge to compare *a priori* approach with *agnostic* analyses that would perform variable selection treating all exposures in the same way, neither answered a large number of causal questions referring to different exposures using causal inference techniques for high-dimensional data. Results from mediation analyses using cross-sectional omics should be interpreted with caution since the omics markers might be a consequence of the health outcome or of the exposure (one or the other, not both). In particular serum metabolome data, which mainly includes information on lipid metabolism, are closely related to phenotypic outcomes such as BMI. Therefore, they should be expected to reflect more outcome classification (e.g. obesity subtypes) than an exposure effect.

We note that this article does not provide an exhaustive list of potential methods to study the exposome. In addition, the provided dataset was limited in power to study exposure interaction, as it is the case for most epidemiological datasets with exposome data. In the future, a potential complementary approach would be the integration of *a priori* knowledge, such as from experimental toxicological data, to find chemical interaction or *a priori* group of exposures with similar targets. This knowledge-driven approach can also be used to reduce the dimensionality in the omics dataset. Future challenge for exposome data should include longitudinal datasets, as it is one of the key components in the exposome definition, the measure of cumulative effects across a lifespan, which was only partially covered here with the data available at two periods, pregnancy and childhood.

The strength of this event was the application of various methods on the same, well-characterized HELIX dataset. Indeed, it was possible for approaches focusing on the same outcome, child BMI, to find the same chemicals as main predictors (PCBs, metals). Although biological interpretation from these analyses should be avoided due to the partially simulated nature of the data and the heterogeneity in the methods to deal with the confounder structure (e.g., multi-center structure), this exposome dataset could serve as a reference to test novel methods in the future. Finally, the work herein has resulted in computational algorithms with associated code made available to the community with an open-source licence, allowing for reproducible research and applications to other similar research questions based on exposome and even multi-omics data. This event was thought-provoking and highlighted the importance of networking between researchers in a multi-disciplinary environment. It fostered future collaborations at a time where interpersonal interaction is constrained by COVID pandemic, as well as giving visibility to researchers of various genders, backgrounds and career stages.

**Author contributions**

Design of the event: LM, XB, JRG

Reviewing and abstract selection (the event committee): LM, XB, JRG, MCH, PMP, CW

Data collection, preparation: XB, CW, LM, JRG

Manuscript preparation: JBG, XB, LM

Data analysis: all the event participants

### Consortia

The Exposome Data Challenge Participant Consortium is listed below in alphabetical order, and author affiliations are available in Table S1.

Alfano Rossella, Basu Sanjib, Benavides Jaime, Broséus Lucile, Brunius Carl, Caceres Alejandro, Carli Matthew, Cazabet Rémy, Chattopadhyay Shounak, Chen Yun Hua, Chillrud Lawrence, Conti David, Gennings Chris, Gouripeddi Ramkiran, Iyer S Hari, Jedynak Paulina, Li Huichu, McGee Glen, Midya

Vishal, Mistry Sejal, Mork S Daniel, Pearce L. John, Peruzzi Michele, Pescador Jimenez Marcia, Reimann Brigitte, Roscoe J. Charlotte, Shen Xiaotao, Stratakis Nikos, Wang Ziyue, Wang Congrong, Wheeler David, Wilson Ander, Wu Qiong, Yu Miao, Zhao Yinqi, Zou Fei, Zugna Daniela.

**Acknowledgements**

The administrative and communication team which did a tremendous job, Rodney Ortiz, Yvette Moya-Angeler and Aleix Cabrera. Congratulations to the winners of the challenge: Xiaotao Shen from Stanford University, [Alejandro Caceres](from) ISGlobal and Glen McGee from the University of Waterloo.

**Funding**

The data for the challenge were issued from a study from the European Community's Seventh Framework Programme (FP7/2007-206) under grant agreement no 308333 (HELIX project) and the H2020-EU.3.1.2. - Preventing Disease Programme under grant agreement no 874583 (ATHLETE project). LMaitre is funded by a Juan de la Cierva-Incorporación fellowship (IJC2018-035394-I) awarded by the Spanish Ministerio de Economía, Industria y Competitividad. ISGlobal and the Exposome hub acknowledges support from the Spanish Ministry of Science and Innovation through the "Centro de Excelencia Severo Ochoa 2019-2023" Program (CEX2018-000806-S), and support from the Generalitat de Catalunya through the CERCA Program. JBGuimbaud was supported by a CIFRE PhD fellowship (#2020/1297) from Meersens. MYu was supported by the grant P30ES023515 (National Institute of Environmental Health Sciences, US).